\documentclass[12pt]{article}
\usepackage{amsmath,amssymb}

\makeatletter
\@addtoreset{equation}{section}
\makeatother

\renewcommand{\theequation}{\thesection.\arabic{equation}}

\newcommand{\tr}{\mathop{\mathrm{tr}}\nolimits}
\newcommand{\res}{\mathop{\mathrm{res}}\limits}

\def \Z{\uparrow}
\def \W{\downarrow}

\title{Finite-Size Corrections to Anomalous Dimensions\\ in N=4 SYM Theory}

\author{Martin L\"ubcke and Konstantin Zarembo\thanks{Also at ITEP, Moscow, Russia}\\ \\
         {\it Department of Theoretical Physics}\\
         {\it Uppsala University, SE-751 08 Uppsala, Sweden}}

\def\N{${\cal N}=4$ }
\def\ads{$\mathrm{AdS}_5\times S^5$}
\def\e{{\,\rm e}\,}

\begin{document}          

\maketitle

\begin{flushright}
	\footnotesize\tt
	\vskip -85mm
	ITEP-TH-23/04\\
	UUITP-12/04
	\vskip 7cm
\end{flushright}

\begin{abstract}
The scaling dimensions of large
operators in \N supersymmetric Yang-Mills theory
 are dual to energies of semiclassical strings in $\mathrm{AdS}_5\times S^5$.
 At one loop, the dimensions of large operators
can be computed with the help of Bethe ansatz and 
can be directly compared to the string energies. 
We study finite-size corrections for Bethe states which should describe
 quantum corrections to energies of extended semiclassical strings.
\end{abstract}

\section{Introduction}

The semiclassical regime of the AdS/CFT correspondence 
\cite{Berenstein:2002jq,Gubser:2002tv}
relates 
classical string theory in $\mathrm{AdS}_5\times S^5$ to perturbative
\N supersymmetric Yang-Mills (SYM) theory. This observation is quite
remarkable and is at first glance counterintuitive,
since the "Planck constant" of the AdS string theory  
$1/\sqrt{\lambda } $ is large at weak 't~Hooft coupling $\lambda=g^2_{\mathrm{YM}}N $
in the SYM.
The key point is that the true parameter of the semiclassical expansion
 for a large class of string states 
is not $1/\sqrt{\lambda } $ but $1/L$ \cite{Tseytlin:2003ii}, where $L$ is a global charge of
the string or some other large quantum number. At a given order in $1/L$,
 $1/\sqrt{\lambda } $ combines with $L$ into the BMN coupling $\lambda /L^2$
\cite{Berenstein:2002jq}
that can be large or small depending on $L$. It is customary now to
identify the limit of small BMN coupling with the perturbative regime in the dual
SYM theory.

The simplest quantities that can  be directly compared to string theory are
scaling dimensions of local operators. They are dual to energies of the
string sates in $\mathrm{AdS}_5\times S^5$. The loop corrections to
scaling dimensions can be readily computed by standard
diagrammatic techniques for simple operators, but to describe semiclassical string states
we need operators with large quantum numbers. Straightforward diagrammatic 
calculations for such operators are hopeless because of
the operator mixing which becomes more and more involved as the
size of the operators grows. Fortunately, the problem 
simplifies considerably in the large-$N$ limit when only planar diagrams contribute
and can be reformulated in a useful way which makes it
tractable. This reformulation also provides a nice physical interpretation
of the operator mixing by  identifying the mixing matrix with a 
Hamiltonian of a certain one-dimensional spin system,
which, quite remarkably, turns out to be
integrable and exactly solvable 
\cite{Minahan:2002ve,Beisert:2003tq,Beisert:2003yb,Beisert:2003ys}.  
The eigenvalues (anomalous dimensions)
can then be calculated using the Bethe ansatz \cite{Bethe:1931hc,Faddeev:1996iy}
and compared to string energies.
This has led to spectacular tests of the AdS/CFT correspondence for
a variety of string solutions with large quantum numbers
\cite{Frolov:2003qc}--\cite{Stefanski:2004cw},\cite{Tseytlin:2003ii}. 

The relationship
between classical strings and operators can be established at one and two loops
quite generally, at the level of effective actions
\cite{Kruczenski:2003gt,Kruczenski:2004kw,Hernandez:2004uw,Stefanski:2004cw} 
or by identifying integrable structures
\cite{Kazakov:2004qf}, which are present both
 in the spin chain  \cite{Faddeev:1996iy}  and
on the string world-sheet 
\cite{Bena:2003wd,Mandal:2002fs,Dolan:2003uh}. 
The purpose of the
present paper is  to go beyond the leading order
in $1/L$ in the SYM,
which on the string side should correspond to taking into account
quantum corrections.
 $1/L$ corrections to the point-like (BMN) 
string states \cite{Berenstein:2002jq} were studied in 
\cite{Parnachev:2002kk,Callan:2003xr,Callan:2004uv}. 
They can be compared to anomalous dimensions of near-BPS operators known
exactly in the first few orders of perturbation theory
\cite{Beisert:2003tq}. 
The situation is more complicated for the extended string solutions. $1/L$ correction
to the string energies are
known for a particular class of solutions
\cite{Frolov:2003tu,Tseytlin:2003ii}, 
but the dual anomalous dimensions have
been so far calculated only in the strict $L= \infty $ limit. 

We shall consider the subsector
of operators of bare dimension $L$ made of two complex scalars: $Z=\Phi _1+i\Phi _2$
and $W=\Phi _3+i\Phi _4$:
$$
\tr ZZZWWZWWZZZZ \ldots
$$
Linear combinations of such operators are in one-to-one correspondence with the states
in the Hilbert space of a periodic spin-1/2 chain, where $Z$ is identified with
spin up and $W$ with spin down on each site of the lattice:
$$
\left| \Z\Z\Z\W\W\Z\W\W\Z\Z\Z\Z \ldots \right\rangle.
$$
The chain is periodic because of the cyclicity of the trace, which also  imposes
the condition of translation invariance (zero-momentum condition) 
on admissible states. The one-loop mixing
matrix in this sector has a form of the Heisenberg Hamiltonian
\cite{Minahan:2002ve}:
\begin{equation}\label{hm}
\Gamma=\frac{\lambda }{16\pi ^2}\sum_{l=1}^{L}
\left(1-\boldsymbol{\sigma }_l\cdot\boldsymbol{\sigma }_{l+1}\right) 
\end{equation}
The higher-loop generalizations of this Hamiltonian,
potentially also  integrable, have been 
discussed in the literature 
\cite{Beisert:2003tq,Beisert:2003jb,Serban:2004jf,Ryzhov:2004nz,Beisert:2004hm}.
The one-loop Hamiltonian can be explicitly diagonalized by Bethe ansatz
\cite{Bethe:1931hc,Faddeev:1996iy}.
The Bethe ansatz works also at two and three loops 
\cite{Serban:2004jf}
and
possibly extends to higher loop orders
\cite{Beisert:2004hm}, 
but we will concentrate on the one-loop anomalous dimensions in this 
paper\footnote{The Bethe ansatz of \cite{Serban:2004jf,Beisert:2004hm}
is only asymptotic and requires $L$ to be large, but corrections to it 
seem to be exponentially small in the large $L$ limit.}.

The eigenstates of the Heisenberg Hamiltonian with $M$ down spins and
with $L$ and $M$ large
are dual to strings localized at the center of AdS and rotating on the five-sphere with
two angular momenta $J_1=L-M$ and $J_2=M$. If the momenta satisfy 
\begin{equation}\label{jc}
m_1J_1+m_2J_2=0,
\end{equation}
where $m_1$ and $m_2$ are integers, 
the classical solutions are very simple \cite{Arutyunov:2003za}. 
The angles on $S^5$ are then linear functions of the world-sheet coordinates. 
We shall concentrate on the gauge theory duals of these simplest 
uniform solutions.

\section{Bethe ansatz}

The vacuum of the Hamiltonian (\ref{hm}) is an empty state
with all spins up and
corresponds to the chiral primary operator $\tr Z^L$. The operators
of the form $\tr( Z^{L-M}W^M+{\rm permutations})$ correspond to the states with
$M$ flipped spins and are characterizes by rapidities $u_i$ of  $M$
magnons. 
The Bethe ansatz
imposes a set of algebraic equations on the rapidities
\cite{Bethe:1931hc,Faddeev:1996iy}:
\begin{equation}
\left(\frac{u_j+i/2}{u_j-i/2}\right)^L=\prod_{k\neq j} \frac{u_j-u_k+i}{u_j-u_k-i}\,.
\end{equation}
The cyclicity of the trace
requires that the total momentum is zero or, in terms of rapidities, that
\begin{equation}\label{mc}
\prod_{j} \frac{u_j+i/2}{u_j-i/2}=1.
\end{equation} 
The anomalous dimension is given by
\begin{equation}
	\label{dimension}
\gamma =\frac{\lambda }{8\pi ^2}\sum_{i}\frac{1}{u_i^2+1/4}\,. 
\end{equation}
These equations completely determine the spectrum of the Hamiltonian
(\ref{hm}).

We are interested in the limit when $L\rightarrow \infty $ and $M\rightarrow \infty $
and the filling fraction $\alpha =M/L$ is kept finite. Some particular
solutions of Bethe equations in this scaling
limit have been discussed in condensed matter
literature \cite{sutherland,AS1} 
and it is this limit that corresponds to the AdS duals 
of the semiclassical strings
\cite{Beisert:2003ea}. 
An inspection of the Bethe equations shows that
 Bethe roots scale with $L$ as $u_i\sim L$. This allows us to
simplify Bethe equations a bit.
Writing them in the logarithmic form,
\begin{equation}\label{logbethe}
L\ln\left(\frac{u_j+i/2}{u_j-i/2}\right)
=\sum_{k\not= j}\ln\left( \frac{u_j-u_k+i}{u_j-u_k-i}\right)-2\pi n,
\end{equation}
and expanding in $1/u_i$, 
we get
\begin{eqnarray}
	\label{betheapprox}
	\frac{L}{u_i} + 2\pi n  =  \sum_{j\neq i} \frac{2}{u_i-u_j}\, .
\end{eqnarray}
An arbitrary phase $2\pi n$ in (\ref{logbethe}) 
reflects the multivaluededness of the logarithm.
In principle, different phases can be assigned to different Bethe roots,
but a macroscopic number of 
roots should have the same phase to get a meaningful scaling
limit at $L\rightarrow \infty $. The
simplest case of equal phases corresponds to the SYM duals
of the uniform string solutions
\cite{Kazakov:2004qf}.
We can indeed easily check that  (\ref{jc}) is satisfied.
Adding together the rescaled Bethe equations
(\ref{betheapprox}), we find the total momentum:
\begin{equation}
P=\sum_{i=1}^{M}\frac{1}{u_i}=-\frac{2\pi nM}{L}, 
\end{equation}
but  (\ref{mc}) implies that the total momentum is 
an integer multiple of $2\pi$, so
\begin{equation}
	\label{fraction}
\alpha \equiv \frac{M}{L}=\frac{m}{n},
\end{equation}
which is equivalent to (\ref{jc}) if we put $m_1=-m$, $m_2=n-m$.

Since Bethe equations are even in $u_i$, the corrections to  (\ref{betheapprox}) are of
order $1/L^2$ and thus  (\ref{betheapprox})  is accurate up to 
and including $O(1/L)$.  In the strict
$L=\infty $ limit, (\ref{betheapprox}) becomes a singular integral equation, which can be 
solved explicitly by the Riemann-Hilbert method
\cite{Kazakov:2004qf}
in the general case when the 
phases are different for different Bethe roots. 
 If all phases are equal, the 
Riemann-Hilbert problem has an algebraic solution. 
As discussed in the appendix, the Riemann-Hilbert
 approach can then be generalized
to take into account the leading $1/L$ correction. Here we use
another method based on the observation of \cite{AS2} 
(essentially present in the earlier work \cite{St,Calogero,Calogero2,Muttalib})
that the {\it exact} solution of  (\ref{betheapprox}) 
can be expressed in terms of the roots of associated Laguerre 
polynomials.

Let us briefly review the arguments of \cite{AS2}. 
Consider the function defined as
\begin{eqnarray}
	Q(u) = \prod_{k=1}^M (u-u_k)
\end{eqnarray}
where $\{u_i\}$ are the roots of (\ref{betheapprox}). This
function
is known as the eigenvalue of the Baxter Q-operator 
\cite{baxter}.
The derivatives of $Q$ evaluated at a root $u_i,$ satisfy
$$u_i Q^{\prime\prime}(u_i)-(L+2\pi n u_i) Q^\prime(u_i)=0$$ in virtue of
 (\ref{betheapprox}), and thus the function 
$u Q^{\prime\prime}(u)-(L+2\pi n u) Q^\prime(u)$ is a polynomial of degree $M$ 
which has the same roots $\{u_i\}$ as $Q(u)$. Hence, this function is
just $Q(u)$, up to a coefficient. Comparing the $u^M$-terms one finds that
\begin{eqnarray}\label{diffQ}
	 u Q^{\prime\prime}(u) - (L+2\pi n u) Q^\prime(u) + 2\pi n M Q(u)=0.
\end{eqnarray}
The polynomial solution of this
 differential equation is the associated Laguerre 
polynomial: $Q(u) \propto  L_M^{-(L+1)}(2\pi n u)$. 
So, the roots of the set of 
equations (\ref{betheapprox}) are the roots of
\begin{eqnarray}
	L_M^{-(L+1)}(2\pi n u_i)=0.
\end{eqnarray}
Some precautions have to be made here since the upper 
index of the associated Laguerre polynomial is negative. 
Using the Rodrigues representation 
and the fact that $L+1>M$ we find:
\begin{eqnarray}\label{lag}
L_M^{-(L+1)}(x) = \frac{\e^x x^{L+1}}{M!} \left( \frac{\mbox d}{\mbox d x} \right)^M e^{-x} x^{M-L-1}
=\sum_{i=0}^M (-1)^M \left( \begin{array}{c} L-i\\ M-i \end{array} \right)	\frac{x^i}{i!}\,.
\end{eqnarray}

We can now use this result to calculate the anomalous dimension in (\ref{dimension}) 
using the relation
\begin{eqnarray}
	\sum_{i=1}^M \frac1{u_i^2}
	 =	\frac1{2\pi i} \oint_\Gamma \frac{\mbox d \omega}{\omega^2} 
\frac{Q^\prime(\omega)}{Q(\omega)}
\end{eqnarray}
where $\Gamma$ is a curve encircling all $\{u_i\}$ counterclockwise. 
We replaced $u_i^2+1/4$ by $u_i^2$, because $u_i=O(L)$ and we 
are interested only in the $1/L$ correction to the anomalous dimension. 
The only singularity outside of the contour of integration is a pole at  $\omega=0$ .
The  contour can be deformed such that the integral picks up the residue:
\begin{equation}
	\sum_i \frac1{u_i^2} 
	 =  -\res_{\omega=0} \left(\frac{Q^\prime(\omega)}{\omega^2 Q(\omega)}\right)
	= -(2\pi n)^2
	\left( \frac{ L(0) L^{\prime\prime}(0) - \left(L^\prime (0)\right)^2}{ \left(L(0)\right)^2} \right) .		
\end{equation}
Using the explicit form of the Laguerre polynomials (\ref{lag}), we find:
\begin{equation}
	\frac{ L(0) L^{\prime\prime}(0) - \left(L^\prime (0)\right)^2 }{ \left(L(0)\right)^2}
		=	\frac{M(M-1)}{L(L-1)} - \frac{M^2}{L^2} = \frac{\alpha(\alpha-1)}{L-1}\,,
\end{equation}
where the definition of the filling fraction (\ref{fraction}) was used. 
Finally, (\ref{dimension}) yields:
\begin{eqnarray}\label{fing}
\gamma
		=	\frac{\lambda n^2 \alpha (1-\alpha)}{2L}\left(1+\frac{1}{L}\right) 
+ O\left( \frac1{L^3} \right)
=\frac{\lambda m(n-m)}{2L}\left(1+\frac{1}{L}\right) + O\left( \frac1{L^3} \right).
\end{eqnarray}
This is our final result.
The finite-size correction to the anomalous dimension 
turns out to be surprisingly simple and is just proportional to the 
leading order. 

\section{Discussion}

We computed the finite-size correction to the energies of the simplest Bethe states that describe
anomalous dimensions of large operators in \N SYM. It would be interesting
 to calculate finite-size corrections to anomalous dimensions for a general Bethe state.
Indeed, the scaling
solution of the Bethe equations is known in complete generality and can 
be expressed in terms of Abelian differentials on hyperelliptic Riemann
surfaces \cite{Kazakov:2004qf}. The resolvent of Bethe roots plays the central role in this construction,
and one may hope that calculations in the appendix can be 
reformulated in a similar geometric language and generalized
to include arbitrary Bethe states. The calculations in the main text
are based on another method, which
reveals an interesting connection to 
the Baxter's Q-operator.  
The eigenvalues of the Q-operator are known to obey a finite difference equation
 equivalent to the set of Bethe equations
\cite{baxter}. On the other hand, the same eigenvalues
 satisfy a linear differential equation (\ref{diffQ}) in the scaling limit.
We were not able to derive
(\ref{diffQ}) directly from the Baxter equation, but it is possible that some relationship 
between them exists. 

The string duals of Bethe states discussed in this paper
are known at the classical level and are described by very simple string configurations
in \ads. It would be very interesting to compute quantum corrections
to their energies and to compare them to the $1/L$ corrections to
anomalous dimensions calculated in this paper. However, the relevant  string solutions
 are unstable \cite{Arutyunov:2003za}, and it is not quite clear how to take into account
quantum corrections for them\footnote{We are grateful to A.~Tseytlin
for a discussion of this point.}. 
This problem is rather puzzling, since the Bethe states at hand seem to be
perfectly well-defined eigenstates of the Heisenberg Hamiltonian and show no
signs of instability.
The resolution of this problem may provide further insights on the relationship between
\N SYM and strings.

\subsection*{Acknowledgments}
We would like to thank V.~Kazakov and A.~Tseytlin for useful suggestions.
The work of K.Z. was supported in part by the Swedish Research Council
under contract 621-2002-3920, by G{\"o}ran Gustafsson Foundation,
and by RFBR grant NSh-1999.2003.2 for the support of scientific schools.

\appendix
\renewcommand{\thesection}{Appendix \Alph{section}:}
\renewcommand{\theequation}{\Alph{section}.\arabic{equation}}

\section{Finite-size corrections from loop equation}

The scaling limit of the Bethe equations (\ref{betheapprox}) has the same form
as the saddle point equation for eigenvalues 
of a Hermitian random matrix
\cite{Brezin:1977sv}.
The analogy becomes complete if we rescale Bethe roots and define $x_i=u_i/L$,
which is finite in the limit of $L\rightarrow \infty $. The rescaled variables satisfy
\begin{eqnarray}\label{ba}
\frac{1}{x_i}+2\pi n=\frac{2}{L}\sum_{j=1}^{M}\frac{1}{x_i-x_j}  
\end{eqnarray}
and 
\begin{eqnarray}\label{momc}
	P=\frac1L \sum_{j=1}^{M} \frac1{x_j} = -2\pi m
\end{eqnarray}
An efficient way to solve matrix models, 
which turns out to be useful also
in the present context, is to reformulate the problem in terms
of the resolvent
\begin{eqnarray}
	G(x) = \frac1L \sum_{j=1}^{M} \frac1{x-x_j}\,.	
\end{eqnarray}
The resolvent can be regarded as a generating function for 
eigenvalues of the commuting Hamiltonians of the Heisenberg model
\cite{Arutyunov:2003rg,Engquist:2003rn}:
\begin{equation}\label{momg}
P=-G(0), \qquad \gamma =-\frac{\lambda }{8\pi^2 L}\,G'(0), \qquad \ldots
\end{equation}
In the matrix models the resolvent satisfies the loop equation
\cite{Migdal:gj,Makeenko:tb}, an analog of which
can be derived by multiplying both sides of (\ref{ba}) by $1/(x-x_i)$ and
summing over $i$. Then
\begin{eqnarray*}
&&
\frac2{L^2} \sum_{i\neq j} \frac1{(x_i-x_j)(x-x_i)}
=\frac1{L^2} \sum_{i\neq j} \frac1{x_i-x_j}
				\left( \frac1{x-x_i} - \frac1{x-x_j} \right)
\\ &&=\frac1{L^2} \sum_{i\neq j} \frac1{(x-x_i)(x-x_j)}
 =
\frac1{L^2} \sum_{ij} \frac1{(x-x_i)(x-x_j)} - \frac1{L^2} \sum_i \frac1{(x-x_i)^2}
\\ &&
=G^2(x) + \frac1{L} G^\prime (x).
\end{eqnarray*}
 Similar manipulations with the left-hand side give:
\begin{eqnarray}\label{er}
	x G^2(x) + \frac1L x G^\prime (x) =  G(x) + 2\pi n x G(x)-2\pi m,
\end{eqnarray}
where the momentum condition (\ref{momc}) in the form (\ref{momg})
was taken into account. We can now expand
 the resolvent in the powers of $1/L$:
\begin{eqnarray}
	G(x) = G_0(x) + \frac1L G_1(x) + O \left( \frac1{L^2} \right)
\end{eqnarray}
and plug this expansion in (\ref{er}):
\begin{eqnarray}
	\label{L0}
	L^0 &:& 0 = x G_0^2 - (1+2\pi nx)G_0 + 2\pi m \\
	\label{L1}
	L^{-1} &:& 0 = 2x G_0 G_1 -(1+2\pi nx)G_1 + x G_0^\prime.
\end{eqnarray}
The leading order is an algebraic equation whose solution is
\begin{eqnarray}
	G_0 = \frac1{2x}\left( 1 + 2\pi nx - \sqrt{(1 + 2\pi nx)^2 - 8\pi n\alpha x} \right),
\end{eqnarray}
where (\ref{fraction}) has been used to express $m$ in terms of the filling
fraction $\alpha $. Alternatively, we
could derive  (\ref{fraction})  from the loop equation
by imposing the boundary condition $G(x)\rightarrow \alpha/x$ at infinity. 
The first correction can be found from equation (\ref{L1}):
\begin{eqnarray}
	\nonumber
	G_1 &=& \frac{x G_0^\prime}{1 + 2\pi nx - 2x G_0}\\ 
			&=& \frac1{2x} \left[ \frac{1+2\pi n(1-2\alpha)x}{(1+2\pi n x)^2-8\pi n\alpha x} 
					- \frac1{\sqrt{(1+2\pi nx)^2-8\pi n \alpha x}} \right].
\end{eqnarray}
Thus, we have the resolvent up to order $L^{-1}$. Its Taylor expansion
\begin{eqnarray}
G(x) &=& 2\pi n \alpha 
+ (2\pi n)^2\alpha(\alpha -1)\left(1+\frac{1}{L}\right)x + \ldots
\end{eqnarray}
and (\ref{momg}) now yield 
$$
\gamma
=	\frac{\lambda n^2 \alpha (1-\alpha)}{2L}\left(1+\frac{1}{L}\right) 
+\ldots\,,
$$
in agreement with (\ref{fing}).

\end{document}